# Magnetization states and switching in narrow-gapped ferromagnetic nanorings

Jie Li[1], Sheng Zhang[1], Chris Grigas[1], Rajiv Misra[1], Jason Bartell[1], Vincent H. Crespi[1], Peter Schiffer[1*]

[1]*Department of Physics and Materials Research Institute, Pennsylvania State University, University Park, PA 16802, USA*

We study permalloy nanorings that are lithographically fabricated with narrow gaps that break the rotational symmetry of the ring while retaining the vortex ground state, using both micromagnetic simulations and magnetic force microscopy (MFM). The vortex chirality in these structures can be readily set with an in-plane magnetic field and easily probed by MFM due to the field associated with the gap, suggesting such rings for possible applications in storage technologies. We find that the gapped ring edge characteristics (i.e., edge profile and gap shape) are critical in determining the magnetization switching field, thus elucidating an essential parameter in the controls of devices that might incorporate such structures.

* Corresponding author: pes12@psu.edu

Ferromagnetic nanorings have long been suggested for use in high-density magnetic storage and random access memories that exploit their stable and reproducible vortex states and flux closure.[1] In the proposed devices, the two chiral magnetization states of magnetic rings represent the binary states (0 or 1). Unfortunately, controllably switching between the states is difficult because of the weak coupling between the closed flux vortex state and external fields. This difficulty can be lifted by introducing asymmetry to the ring structure; chirality control has been demonstrated for asymmetric,[2-5] elliptical,[6-8] square[4, 9] and triangular[10] rings. The closed magnetic flux in symmetric rings also impedes probing of the chirality of vortex states, although this problem can be alleviated with electrical measurements[11-12] or by pac-man or slotted-ring shapes with geometric restrictions[13-16] that allow flux leakage.

Because they retain the flux states of complete rings and yet are more readily controllable, the asymmetric variants on ferromagnetic nanorings have tremendous potential for technological applications. However, many asymmetric ring variants either are closed loops (which are difficult to probe magnetically with standard magnetoresistive sensors) or have considerable flux leakage in the plane (which would produce strong magnetostatic interactions in dense arrays). One structure that is subject to neither of these concerns is a ferromagnetic nanoring with a narrow gap that breaks the ring symmetry.[13-15] Such a gap preserves the vortex state, but also allows the chirality of the state to be easily probed due to the magnetic flux within the gap, even though the stray field elsewhere in the plane of the ring is minimal as reflected by the micromagnetic calculations (as it is for fully symmetric rings). We demonstrate that the chiral state of such gapped rings can be well-controlled through application and removal of a saturating field along a symmetry-breaking axis. Through micromagnetic modeling, we find that the detailed switching behavior of the gapped rings is determined by a combination of the edge roughness and curvature of the gap edge structure. These findings demonstrate the importance of edge details on the behavior of ferromagnetic nanoring structures, and they will also enable more complex device designs utilizing narrow-gapped ferromagnetic rings.

We lithographically fabricated the isolated permalloy ($Ni_{0.81}Fe_{0.19}$) nanorings by electron-beam lithography following procedures published previously.[17-19] The gapped nanorings were designed with inner and outer radii of 200 nm and 300 nm respectively, and a gap with a subtended width of ~ 20 degrees. A 20 nm thick polycrystalline permalloy film was deposited onto the resist patterns with molecular beam epitaxy (MBE) followed by liftoff. These polycrystalline films of near-zero magnetocrystalline anisotropy enabled us to freely control the magnetization states by designing the geometry of the nanostructures. The fabricated nanorings, as shown in the SEM image [Fig. 1(b)], have rounded ends in the gap and rougher edges compared to the ideal designed shape [Fig. 1(a)] due to limitations in the nanofabrication process. Such roughness, typical of the limitations of e-beam lithography, is apparent in all published experimental studies of ferromagnetic nanostructures,[2, 15-16, 20-23] although we are not aware of any previous study of its impact on the global magnetic properties of ring structures.

The magnetic moments of gapped nanorings are constrained in the **x-y** plane due to strong shape anisotropy. We simulate the magnetic behavior of the rings using the OOMMF package,[24] assuming $5 \times 5 \times 5$ $nm^3$ for the elements of the simulation,[25] and image them directly by MFM (Digital Instruments NanoScope IV) using low-moment and regular-moment Veeco tips (MESP-LM and MESP) at a scan and lift height of 50 nm. The dark regions in the MFM images indicate a north pole of the structure while the light regions indicate a south pole [Figs. 1(c-d) at low resolution and Fig.2 at higher resolution].

Analogous to full symmetric nanorings, the lowest energy configuration of gapped nanorings is a vortex state circling the ring.[5, 21] We initialize the rings in the vortex state following our previously developed protocol of ac demagnetization – the samples were rotated in-plane while subjected to a stepwise decreasing in-plane external magnetic field.[17-19, 26] Unlike the complete rings, however, the vortex state of gapped rings is readily accessible to local magnetic imaging because of the strong localized field within the gap (the simulations indicate that the field at the center point of the gap can be as high as 925 Oe at remanence and drops rapidly to less than 100 Oe within 85 nanometers outside the gap). The vortex state has two configurations of different chirality, clockwise and counter-clockwise, as shown by MFM images in Figs. 1(c-d). Over multiple runs on different locations on 140 rings, the vortex chirality can be set with 95% certainty by applying and removing a sufficiently large field to saturate the moment in the **y** direction, thus breaking inversion symmetry as has been shown for other asymmetric structures.[16] The gap is necessary for this control, since the ungapped ring under external field cannot distinguish clockwise and counter-clockwise states on symmetry grounds.

We now examine the magnetization of the gapped rings upon application of fields along **x** and **y**, starting from the vortex state. In both cases, the magnetization increases sharply at a characteristic field, $H_c$ when a slightly polarized vortex state transitions to an almost completely polarized onion state analogous to that found in complete nanorings.[7, 21-23, 27-28] Fig. 2 compares MFM images with OOMMF-simulated magnetization configurations of gapped nanorings as the external magnetic field increases along **x** (top panel) and **y** (bottom panel) for $H = 0$, $H < H_c$, $H \gtrsim H_c$ and $H >> H_c$ (i.e. a fully saturated state). The experimental images for both directions were obtained by scanning nanorings starting from the vortex state till the full polarization. When immediately approaching the $H_c$ (80 Oe below $H_c$), a finer step size (16 Oe) was used until the nanorings probed were all switched. Large steps (32 Oe or larger) were used out of this range where the magnetization did not show substantial difference from the initial vortex state or were already switched by the external field (which continues to polarize until near saturation). Both $H < H_c$ states in Fig. 2 retain the overall topology of the vortex state, with some bias due to the weak external field. At low field these deviations are most pronounced near the gap edge and (for the field along **x**) along the top half of the ring. Magnetization reversal for the field along **y** originates at the edges of the gap, while it occurs primarily along the top half of the ring for the field along **x** (the moment along the bottom half is already aligned with the field along **x** even in the vortex state).

In addition to the qualitative comparison in Fig. 2, we can compare the value of $H_c$ determined from our simulations with the measured values determined from in-field MFM scans with a field step size of 16 Oe near the switching. As shown in Fig. 3a, where the red and blue arrows indicate the experimental results, the agreement is excellent and well within our experimental uncertainty (about 16 Oe, indicated by the width of the green lines). Data taken for the field applied in the negative **x** direction did not show any noticeable difference, as would be expected by symmetry considerations. We did not make a detailed study for a magnetic field in the negative **y** direction, i.e., with the field reversing the polarization of the ungapped half of the ring. Low-resolution data for this case, however, do indicate that $H_c$ is higher (665 ± 80 Oe) than for the field applied in the positive **y** direction, in rough agreement with the simulation value of 722.5 Oe.

Since edge roughness in the magnetic nanostructures can affect magnetization switching due to the complexities of local domain structures, the simulations discussed above were based on an actual ring shape obtained from an SEM image. To evaluate the effect of the edge roughness and shape on magnetization switching, we also performed simulations of ideal rings based on the ideal parameters used in the lithography (i.e. the shape shown in Fig 1(a)). As seen in Fig. 3(a), the $H_c$ of ideal gapped rings is higher by 100 Oe and 155 Oe for the **y** and **x** directions respectively; these values are well separated from the experimental results. For comparison, we also show the simulation results for a complete ring with the ideal designed shape, for which $H_c$ is even higher.

The simulation results demonstrate the importance of the rounded gap edges and the edge roughness in determining $H_c$. To differentiate between these two factors, we separately simulated a shape with only a rough edge or only rounded ends. Fig. 3(b) shows the gap outlines and values of $H_c$ for each shape and field direction. The difference is substantial; by independently controlling edge roughness and gap rounding, the ratio of the critical fields in the **x** and **y** directions varies from 1.2 to 1.8, for nanorings of nominally identical overall size and shape. Interestingly, the suppression of the switching field is governed by different mechanisms for **x** and **y** directions. The shape with just a rough edge reduces $H_c^x$ while $H_c^y$ remains essentially the same as that of the ideal shape. By contrast, the rounded ends reduce $H_c^y$ while $H_c^x$ remains essentially the same. Based on the images in Fig 2, we deduce that gap roughness supplies nucleation sites for the magnetization to switch along the full width of the ring and thus lowers the switching field along the **x** direction. By contrast, switching along the **y** direction requires the coherent rotation of spins near the gap edges, which is facilitated by rounded ends.

In conclusion, our results confirm that narrow-gapped rings have two notable advantages over full symmetric rings: their magnetic vortex state can be set by applying an external field perpendicular to the gap axis and the chirality of their vortex state can be easily probed by MFM and other field-sensitive local techniques. Importantly, our results also indicate that the details of the gap and edges are particularly important in determining the overall behavior of the gapped rings, suggesting their use as an effective control parameter through which device properties could be tuned. Future studies could also probe more creative edge shapes, i.e. with carefully controlled roughness or narrower gaps (through optimized layout design and by using higher resolution e-beam systems), as well as devices that exploit the strong and highly localized field within the gap region itself as demonstrated by our micromagnetic simulations.

**Acknowledgements:** This research was supported by the U.S. Department of Energy, Office of Basic Energy Sciences, Materials Sciences and Engineering Division under Award # DE-SC0005313, and by the Army Research Office, the National Nanotechnology Infrastructure Network and an REU Supplement to NSF grant DMR-0701582. We are grateful to Chris Leighton and Mike Erickson for assistance with the sample preparation and to them, Paul Crowell, and Paul Lammert for helpful discussions.

**FIGURE CAPTIONS**

Figure 1: (a) Ideal as-designed outline of the gapped nanoring; the inner radius and width of the nanoring are 200 nm and 100 nm respectively and the gap is about 20 degrees. (b) SEM image of one gapped nanoring; the scale bar is 100 nm. MFM images of the remanent states after the application of magnetic field along the (c) positive and (d) negative **y** axis.

Figure 2: Magnetization states imaged by MFM (top rows) and compared with simulated states (bottom rows) when an increasing external field is applied along (a) the **x** direction where the four states are imaged at 0, 648, 665, and 913 Oe; (b) along the **y** direction where the four states are imaged at 0, 166, 494, and 912 Oe. The zero-field images show the vortex state in both cases. Both measurements and simulations show only two distinct states: a vortex state at remanence ($H \sim 0$) and an onion state after switching ($H \gtrsim H_c$); the states below $H_c$ maintain the vortex topology by forming a deformed circular loop.

Figure 3: (a) Micromagnetic simulations of switching for a gapped nanoring, using the designed shape (dashed lines) and SEM images (solid lines). The SEM image shape shows a much smaller switching field than the ideal shape, in good agreement with the MFM measurements indicated by the arrows (the width of green lines represents a measurement uncertainty of 16 Oe). (b) Comparison of the simulated switching fields for the designed shape and SEM-derived shape, as well as shapes with only rough edge or rounded ends. The last two simulations show the distinct roles of roughness and roundness in **x** and **y** switching.

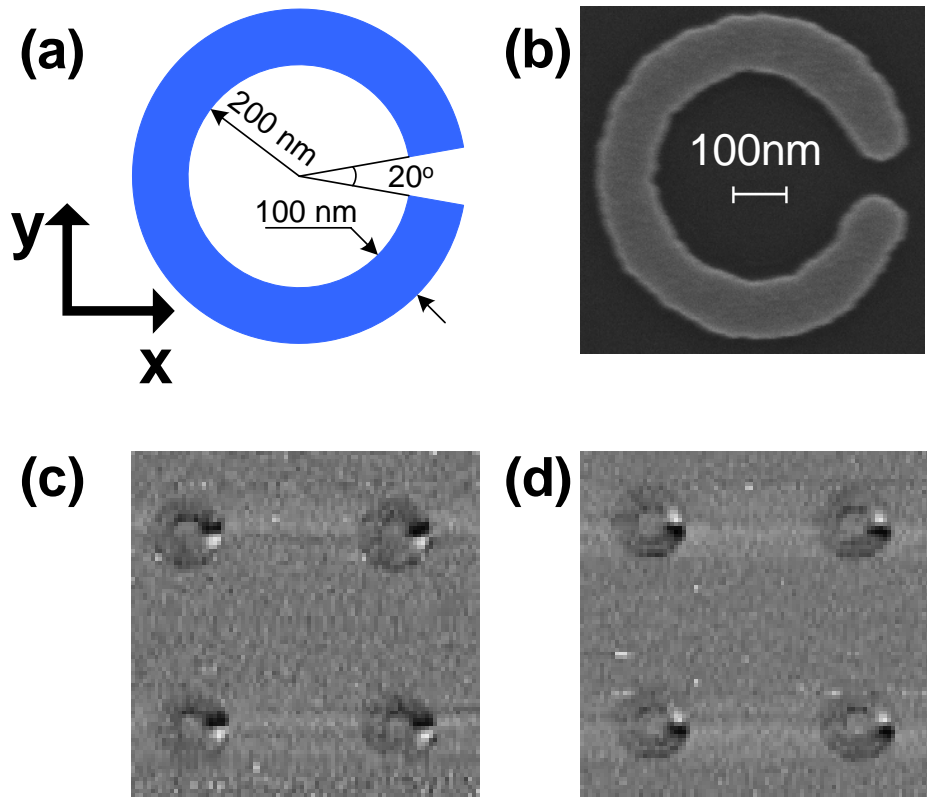

Figure 2

J. Li, *et al.*

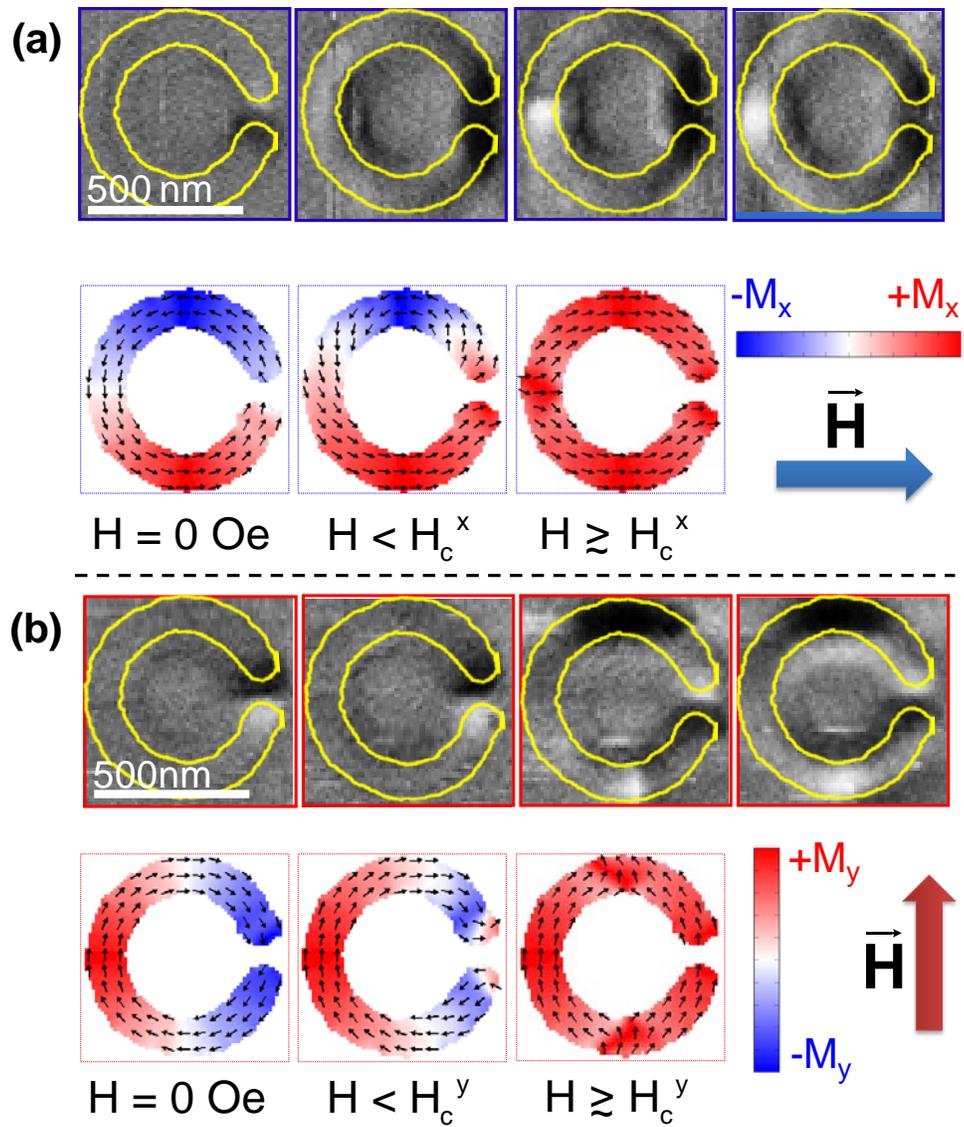

Figure 3(a)

J. Li, *et al.*

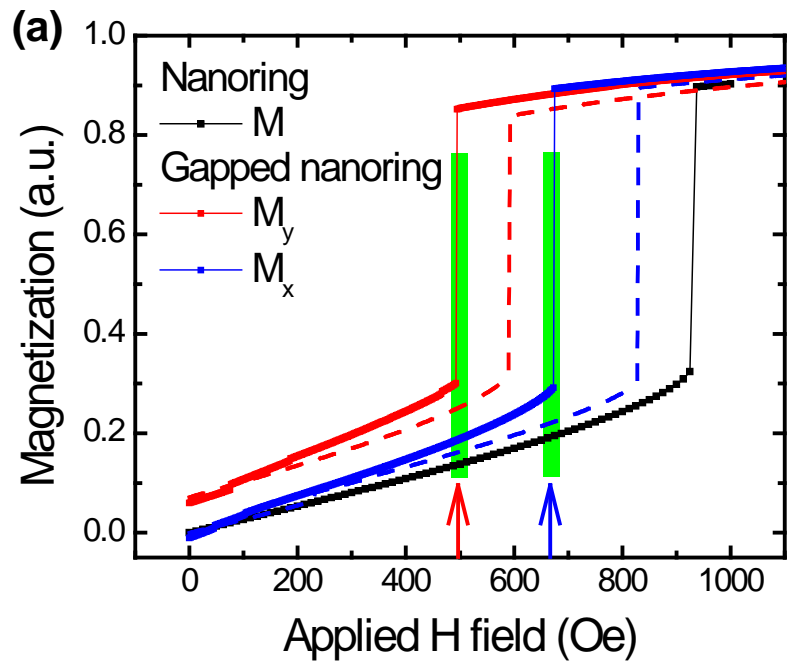

Figure 3(b)

J. Li, *et al.*

| | Outlines | $H_c^x$ (Oe) | $H_c^y$ (Oe) |
|---|---|---|---|
| Measured | | 665 | 494 |
| SEM | 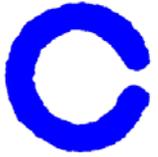 | 675 | 495 |
| Ideal outline | 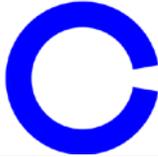 | 830 | 593 |
| Rough edge | 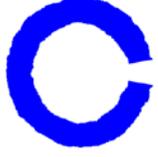 | 738 | 600 |
| Rounded ends | 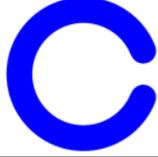 | 838 | 460 |

# References

[1] J. Zhu, Y. Zheng, and G. A. Prinz, J. Appl. Phys. **87**, 6668 (2000).

[2] F. Q. Zhu, G. W. Chern, O. Tchernyshyov, X. C. Zhu, J. G. Zhu, and C. L. Chien, Phys. Rev. Lett. **96**, 027205 (2006).

[3] F. Giesen, J. Podbielski, B. Botters, and D. Grundler, Phys. Rev. B **75**, 184428 (2007).

[4] M. Konoto, T. Yamada, K. Koike, H. Akoh, T. Arima, and Y. Tokura, J. Appl. Phys. **103**, 023904 (2008).

[5] M. Klaui, J. Rothman, L. Lopez-Diaz, C. A. F. Vaz, J. A. C. Bland, and Z. Cui, Appl. Phys. Lett. **78**, 3268 (2001).

[6] W. Jung, F. J. Castano, and C. A. Ross, Phys. Rev. Lett. **97**, 247209 (2006).

[7] C. Nam, M. D. Mascaro, and C. A. Ross, Appl. Phys. Lett. **97**, 012505 (2010).

[8] F. J. Castaño, C. A. Ross, and A. Eilez, J. Phys. D: Appl. Phys. **36**, 2031 (2003).

[9] A. Libal, M. Grimsditch, V. Metlushko, P. Vavassori, and B. Janko, J. Appl. Phys. **98**, 083904 (2005).

[10] A. Westphalen, A. Schumann, A. Remhof, H. Zabel, T. Last, and U. Kunze, Phys. Rev. B **74**, 104417 (2006).

[11] C. Nam, M. D. Mascaro, B. G. Ng, and C. A. Ross, J. Phys. D: Appl. Phys. **42**, 222002 (2009).

[12] C. Nam, B. G. Ng, F. J. Castano, M. D. Mascaro, and C. A. Ross, Appl. Phys. Lett. **94**, 082501 (2009).

[13] M. H. Park, Y. K. Hong, S. H. Gee, D. W. Erickson, and B. C. Choi, Appl. Phys. Lett. **83**, 329 (2003).

[14] H. Hu, H. Wang, M. R. McCartney, and D. J. Smith, Phys. Rev. B **73**, 153401 (2006).

[15] N. Agarwal, D. J. Smith, and M. R. McCartney, J. Appl. Phys. **102**, 023911 (2007).

[16] K. He, N. Agarwal, D. J. Smith, and M. R. McCartney, Magnetics, IEEE Transactions on **45**, 3885 (2009).

[17] R. F. Wang, C. Nisoli, R. S. Freitas, J. Li, W. McConville, B. J. Cooley, M. S. Lund, N. Samarth, C. Leighton, V. H. Crespi, and P. Schiffer, Nature **439**, 303 (2006).

[18] X. Ke, J. Li, C. Nisoli, P. E. Lammert, W. McConville, R. F. Wang, V. H. Crespi, and P. Schiffer, Phys. Rev. Lett. **101**, 037205 (2008).

[19] J. Li, X. Ke, S. Zhang, D. Garand, C. Nisoli, P. Lammert, V. H. Crespi, and P. Schiffer, Phys. Rev. B **81**, 092406 (2010).

[20] M. T. Bryan, D. Atkinson, and R. P. Cowburn, Appl. Phys. Lett. **85**, 3510 (2004).

[21] S. P. Li, D. Peyrade, M. Natali, A. Lebib, Y. Chen, U. Ebels, L. D. Buda, and K. Ounadjela, Phys. Rev. Lett. **86**, 1102 (2001).

[22] F. J. Castaño, C. A. Ross, C. Frandsen, A. Eilez, D. Gil, H. I. Smith, M. Redjdal, and F. B. Humphrey, Phys. Rev. B **67**, 184425 (2003).

[23] J. Rothman, M. Kläui, L. Lopez-Diaz, C. A. F. Vaz, A. Bleloch, J. A. C. Bland, Z. Cui, and R. Speaks, Phys. Rev. Lett. **86**, 1098 (2001).

[24] OOMMF NIST code. http://math.nist.gov/oommf, (2005).

[25] M. A. Akhter, D. J. Mapps, Y. Q. M. Tan, A. Petford-Long, and R. Doole, J. Appl. Phys. **81**, 4122 (1997).

[26] R. F. Wang, J. Li, W. McConville, C. Nisoli, X. Ke, J. W. Freeland, V. Rose, M. Grimsditch, P. Lammert, V. H. Crespi, and P. Schiffer, J. Appl. Phys **101**, 09J104 (2007).

[27] G. D. ChavesO'Flynn, K. Xiao, D. L. Stein, and A. D. Kent, J. Appl. Phys. **103**, 07D917 (2008).

[28] M. Klaui, C. A. F. Vaz, J. A. C. Bland, W. Wernsdorfer, G. Faini, and E. Cambril, Appl. Phys. Lett. **81**, 108 (2002).